\documentclass{moriond}

\def\be{\begin{equation}}
\def\ee{\end{equation}}
\def\bea{\begin{eqnarray}}
\def\eea{\end{eqnarray}}

\newcommand{\tb}{\bar t}

\newcommand{\ttbh}{ t \tb H}
\newcommand{\ttbz}{ t \tb Z}
\newcommand{\ttbw}{ t \tb W}
\newcommand{\ttbb}{ t \tb B}

\newcommand{\als}{\alpha_{\rm s}}
\newcommand{\shat}{\hat s}

\newcommand{\muf}{\mu_{\rm F}}
\newcommand{\mur}{\mu_{\rm R}}
\newcommand{\mufo}{\mu_{{\rm F},0}}
\newcommand{\muro}{\mu_{{\rm R},0}}
\newcommand{\sigh}{\hat \sigma}

\newcommand{\nn}{\nonumber}
\newcommand{\tosv}{{\scriptscriptstyle \to}}

\newcommand{\Photo}{}

\begin{document}
\vspace*{4cm}
\title{ASSOCIATED TOP-PAIR PRODUCTION WITH A HEAVY BOSON PRODUCTION THROUGH NLO+NNLL ACCURACY AT THE LHC}

\author{A. Kulesza }
\address{Institute for Theoretical Physics, WWU M\"unster, D-48149 M\"unster, Germany}
\author{L. Motyka}
\address{Institute of Physics, Jagiellonian University, S.\L{}ojasiewicza 11, 30-348 Krak\'ow, Poland}
\author{D. Schwartl\"ander}
\address{Institute for Theoretical Physics, WWU M\"unster, D-48149 M\"unster, Germany}
\author{T. Stebel}
\address{Institute of Nuclear Physics PAN, Radzikowskiego 152, 31-342 Krak\'ow, Poland}
\author{V. Theeuwes}
\address{Institute for Theoretical Physics, Georg-August-Univesity G\"ottingen, Friedrich-Hund-Platz 1, 37077 G\"ottingen, Germany}

\maketitle\abstracts{In this talk we present results of our recent calculations of cross sections and distributions for the associate production of top-antitop quark pairs with a heavy boson (Higgs, $W$,$Z$) at the LHC,  obtained using threshold resummation in direct QCD, i.e.\ in the Mellin-space approach.}


The measurements\cite{Aaboud:2018urx,Sirunyan:2018hoz,CMS:2019nos} of associated production of a Higgs or a heavy electroweak (EW) boson ($H$,$W$,$Z$) with a top-antitop quark pair provide an important test for the Standard Model at the LHC, in particular
 the top quark couplings. For example, the associated $t \bar t  H$ production  directly probes the top Yukawa coupling without making any assumptions on its nature. 
Fixed order cross sections up to next-to-leading order in $\als$ have been known for some time both for the asociated Higgs boson\cite{Beenakker:2001rj,Reina:2001sf} and $W$ and $Z$ boson production \cite{Lazopoulos:2008de,Lazopoulos:2007bv}. They were recalculated and matched to parton showers in \cite{Hirschi:2011pa,Frederix:2011zi,Garzelli:2011vp,Kardos:2011na,Campbell:2012dh,Alwall:2014hca,Hartanto:2015uka}. Furthermore, next-to-leading order (NLO) QCD-EW corrections are also known \cite{Frixione:2014qaa,Yu:2014cka}.
For the $t \bar t H$ process, the NLO EW and QCD corrections to production with off-shell top quarks were also obtained~\cite{Denner:2015yca,Denner:2016wet}. While next-to-next-to-leading order (NNLO) calculations for this particular type of 2 to 3 processes are currently out of reach, a class of corrections beyond NLO from the emission of soft and/or collinear gluons can be taken into account with the help of resummation methods. Such methods allow to account for effects of soft gluon emission to all orders in perturbation theory. Two  approaches to perform soft gluon resummation are either a direct calculation in QCD or in an effective field theory, in this case soft-collinear effective theory (SCET). 

For the associated $\ttbh$  production, the first calculations of the resummed cross section at the next-to-leading logarithmic (NLL) acurracy, matched to the NLO result were presented in~\cite{Kulesza:2015vda}. The calculation relied on application of the traditional Mellin-space resummation formalism in the absolute threshold limit, i.e.\ in the limit of the partonic energy $\sqrt{\shat}$ approaching the production threshold $M=2 m_t + m_H$. Subsequently,  resummation of NLL corrections arising in the limit of $\sqrt{\shat}$ approaching the invariant mass threshold Q, with $Q^2= (p_t +p_{\bar t}+ p_H)^2$, was performed in~\cite{Kulesza:2016vnq} and later extended to the next-to-next-to-leading-logarithmic (NNLL) accuracy \cite{Kulesza:2017ukk}.  Calculations in the SCET framework  for the $t\bar{t}H$ process led first to  obtaining approximate NNLO~\cite{Broggio:2015lya} and later full NLO+NNLL predictions~\cite{Broggio:2016lfj}.  SCET results at the NLO+ NNLL accuracy for  the $t \bar t W/Z$  production have been also obtained~\cite{Li:2014ula,Broggio:2016zgg}.

In this proceedings we report on our predictions for the threshold-resummed cross sections $pp \rightarrow t \bar t B$, $B=H,W,Z$, obtained using the Mellin-space approach at the NNLL accuracy. 
We treat the soft gluon corrections in the invariant mass kinematics, i.e.\ we consider the limit $\hat \rho = \frac{Q^2}{\shat} \rightarrow 1$ with $Q^2=(p_t + p_{\bar t} + p_{B})^2$ and $\shat$ the partonic center of mass energy. The resummation of large logarithms of $1-\hat \rho$ takes place in the space of Mellin moments $N$, taken w.r.t.\ $\hat \rho$.
At the NNLL accuracy, our key formula for the resummed cross section for the partonic process $ij \tosv \ttbb$  reads\cite{Contopanagos:1996nh,Kidonakis:1998bk,KS2,Bonciani:1998vc,Czakon:2009zw}
\begin{eqnarray}
\label{eq:res:fact_diag}
&\frac{d\tilde{\sigh}^{\rm (NNLL)}_{ij\tosv \ttbb}}{dQ^2} &\!\!\!(N, Q^2,\{m^2\},\muf^2,\mur^2) = {\mathrm{Tr}}\left[ \mathbf{H}_R (Q^2, \{m^2\},\muf^2, \mur^2)\mathbf{\bar{U}}_R(N+1, Q^2,\{m^2\}, Q^2 )  \right. \nn
\\ 
&&\times \left. \mathbf{\tilde S}_R (N+1, Q^2, \{m^2\})\, \mathbf{{U}}_R(N+1, Q^2,\{m^2\}, Q^2 )  \right] \nonumber \\
&&\times\,
\Delta^i(N+1, Q^2,\muf^2, \mur^2 ) \Delta^j(N+1, Q^2,\muf^2, \mur^2 ),\nn
\end{eqnarray}
where $\{m^2\}$ denotes all masses entering the calculations and $\muf$, $\mur$ are the factorization and renormalization scales. The jet functions $\Delta_i$ and $\Delta_j$ account for (soft-)collinear logarithmic contributions  from the initial state partons and are well known at NNLL \cite{Catani:2003zt}. 
$\mathbf{H}_R$, $\mathbf{\bar{U}}_R$,  $\mathbf{{U}}_R$ and $\mathbf{S}_R$ are matrices in colour space over which the trace is taken. The term $\mathbf{\bar{U}}_R\mathbf{\tilde S}_R\mathbf{{U}}_R$ originates from a solution of the renormalization group equation for the soft function and consists of the evolution matrices $\mathbf{\bar{U}}_R$, $\mathbf{{U}}_R$, as well as the function $\mathbf{\tilde S}_R$ which plays the role of a boundary condition of the renormalization group equation. In general the evolution matrices are given by path-ordered exponentials of the soft anomalous dimension matrix $\mathbf{\Gamma}_{ij\tosv \ttbb}(\als)= \left(\frac{\als}{\pi}\right) \mathbf{\Gamma}^{(1)}_{ij\tosv \ttbb} +\left(\frac{\als}{\pi}\right)^2 \mathbf{\Gamma}^{(2)}_{ij\tosv \ttbb}+\ldots$. At NLL, the path-ordered exponentials collapse to standard exponential factors in the colour space $\mathbf R$ where $\mathbf \Gamma^{(1)}_R$  is diagonal. At NNLL, the  path-ordered exponentials are eliminated by treating $\mathbf{U}_R$ and  $\mathbf{\bar{U}}_R$  perturbatively \cite{Buras:1979yt,Ahrens:2010zv}. The  function $\mathbf{H}_R$ accounts for the hard scattering contributions projected on the $\mathbf R$ color basis. At NNLL, the ${\cal O}(\als)$ terms in the perturbative expansion of $\mathbf{H}_R$ and $\mathbf{\tilde S}_R$, as well as  $\mathbf{\Gamma}^{(2)}_R$ are needed. While the latter is known~\cite{Ferroglia:2009ep}, the virtual corrections which enter $\mathbf{H}^{\mathrm{(1)}}_{R}$ are extracted numerically from the NLO calculations provided by {\tt PowHel} \cite{Garzelli:2011vp,Kardos:2011na} and {\tt MadGraph5\_aMC@NLO}~\cite{Alwall:2014hca}.  For more information on the theoretical framework, we refer the reader to our earlier publications \cite{Kulesza:2017ukk}.

The  results for the resummed cross sections are matched the NLO cross sections calculated with {\tt MadGraph5\_aMC@NLO}~\cite{Alwall:2014hca}. In the numerical calculations we use the PDF4LHC15\_30 parton distribution function sets~\cite{Butterworth:2015oua}  and the same input parameters as in the HXSWG Yellow Report~4~\cite{deFlorian:2016spz},  i.e.\ $m_H=125$ GeV, $m_t=172.5$ GeV, $m_W=80.385$ GeV, $m_Z=91.188$ GeV, $G_F=1.1663787\times 10^{-5}$ GeV$^{-2}$, so that we reproduce the NLO values of the $\ttbb$ cross sections listed there. The NNLO sets are employed for the NLO+NNLL predictions, whereas the NLO+NLL predictions are calculated with NLO sets. 

\begin{figure}
\begin{minipage}{0.32\linewidth}
\centerline{\includegraphics[width=0.9\linewidth]{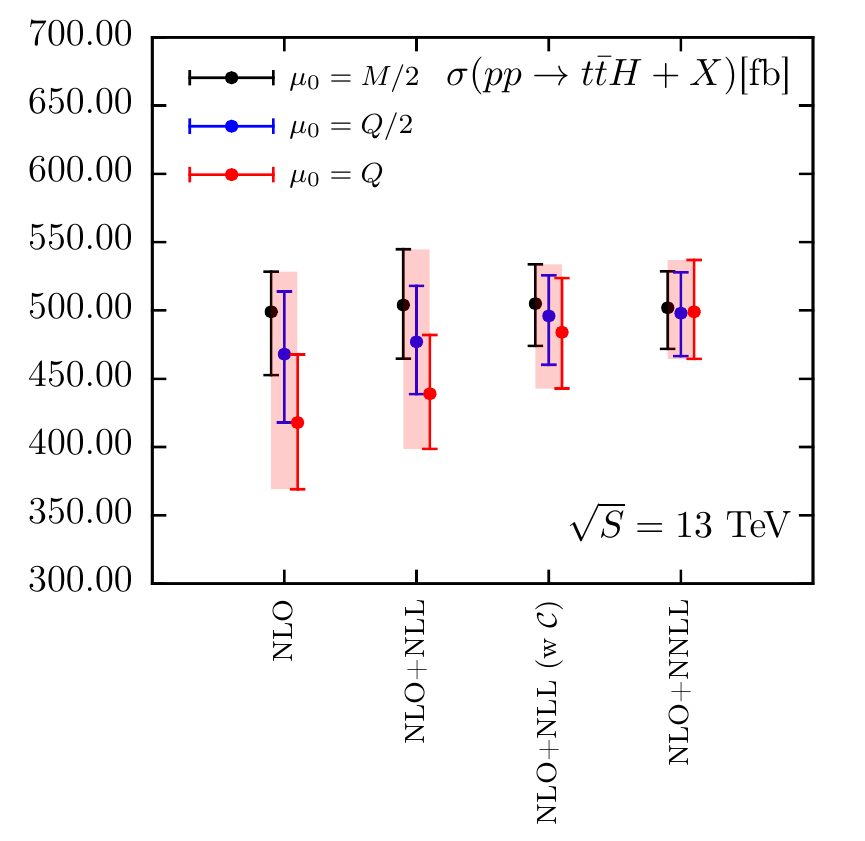}}
\end{minipage}
\hfill
\begin{minipage}{0.32\linewidth}
\centerline{\includegraphics[width=0.9\linewidth]{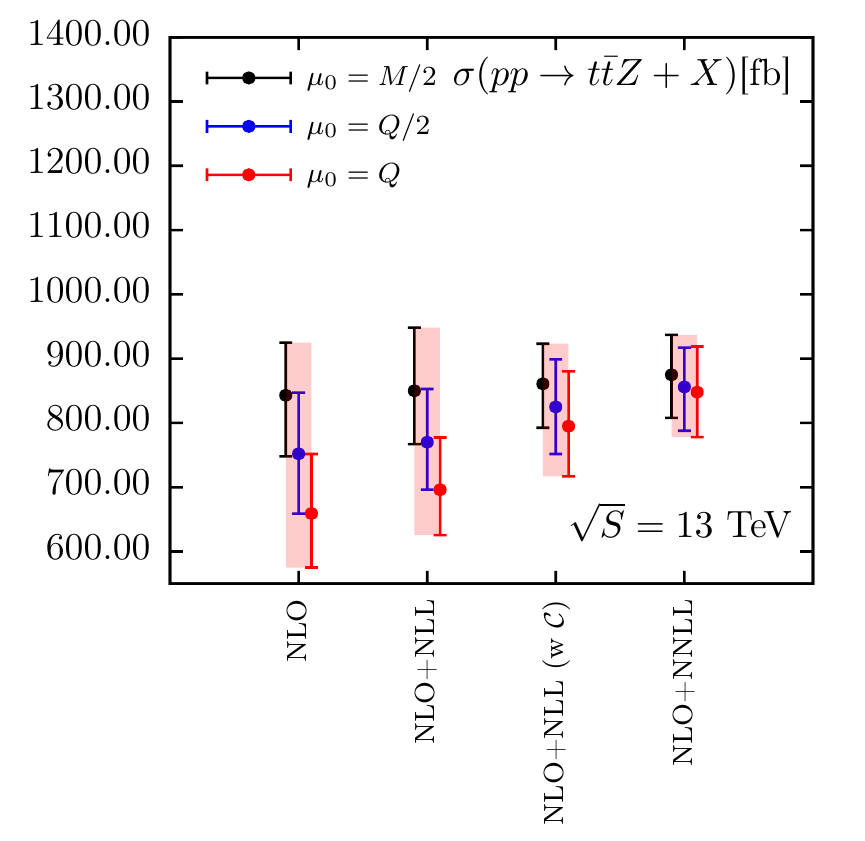}}
\end{minipage}
\hfill
\begin{minipage}{0.32\linewidth}
\centerline{\includegraphics[width=0.9\linewidth]{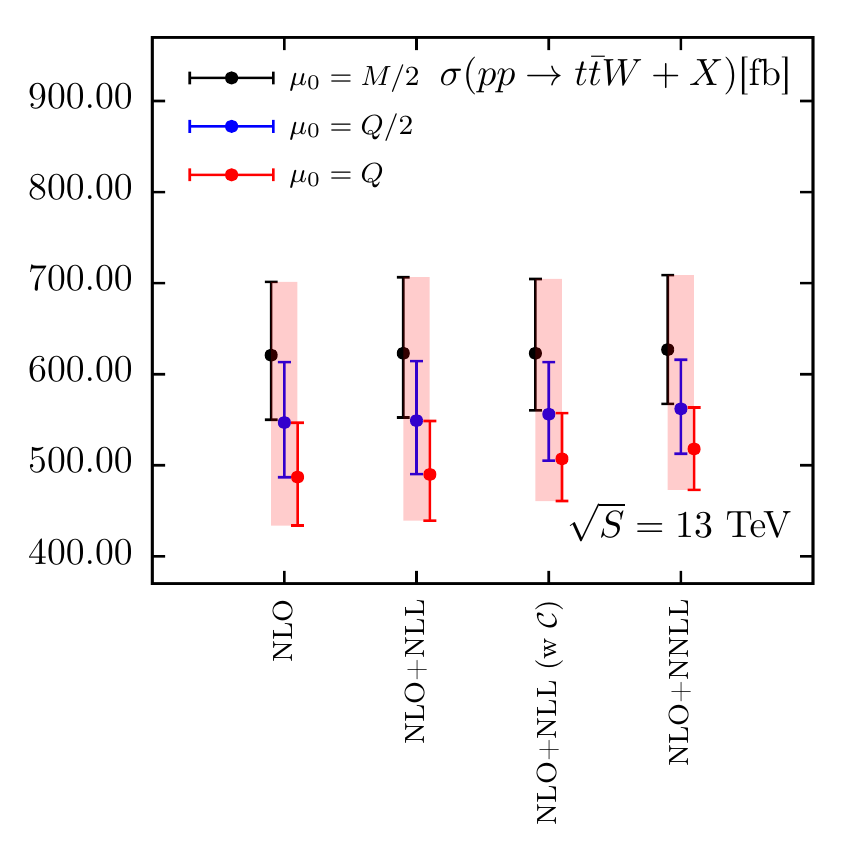}}
\end{minipage}
\caption[]{Total cross sections for the $\ttbh$ (left), $\ttbz$ (center) and $\ttbw$ (right) production at $\sqrt S =13$ TeV, as described in text.} 
\label{fig:totalxsecs}
\end{figure}

In Fig.\ \ref{fig:totalxsecs} we show numerical predictions for the total cross sections  at 13 TeV with three choices  of the central value of the renormalization and factorization scales, $\mu_0=\mufo=\muro$, i.e.\ $\mu_0=Q$, $\mu_0=M/2=m_t+m_B/2$ and  $\mu_0=Q/2$. The theoretical error due to scale variation is calculated using the so called 7-point method, where the minimum and maximum values obtained with $(\muf/\mu_{0}, \mur/\mu_{0}) = (0.5,0.5), (0.5,1), (1,0.5), (1,1), (1,2), (2,1), (2,2)$ are considered. Total cross section results were obtained by integrating the resummed differential cross section. Apart from results at the NLO+NNLL accuracy, we also shown predictions at lower logarithmic accuracy i.e.\ at NLO+NLL and NLO+NLL w ${\cal C}$ as defined and discussed in our latest work~\cite{Kulesza:2017ukk}. Compared to NLO, the NLO+NNLL results demonstrate remarkable stability w.r.t.\ the scale choice, indicating the importance of resummed calculations.   The stability increases  as the accuracy of resummation improves from NLL to NNLL.  In general, resummation leads to positive corrections, bringing the theoretical predictions for the $\ttbz$ and, to a lesser extent, $\ttbw$ total cross sections closer to experimental measurements~\cite{Kulesza:2017ukk}. The relative size of the NNLL corrections w.r.t.\ NLO results differs from 1\% to 19\% (for $\ttbh$) or 4\% to 24\%  (for $\ttbz$), depending on the scale choice. All the trends discussed here are much stronger for the $\ttbh$ and $\ttbz$  than for the $\ttbw$ production due to the $gg$ channel contributing to the LO and, correspondingly, to the resummed cross section.  Another feature of the resummed predictions is a decrease of the scale uncertainties calculated for each specific scale choice, which is also progressing with increasing precision of the theoretical predictions. For example, for the $\ttbh/Z$  production and $\mu_0=Q$ scale choice, the relative size of the scale error is reduced by about 40\%~\cite{Kulesza:2017ukk}. For the other two scale choices the effect is smaller but still sizeable, bringing the value of the scale error down to 5--7\% and 7--8\% for the  $\ttbh$ and $\ttbz$ production, respectively. Thus, while at NLO the accuracy of the $\ttbz$  predictions is worse than the experimental precision, the accuracy of the NLO+NNLL calculations matches the latest experimental precision~\cite{CMS:2019nos}.

\begin{figure}
\begin{minipage}{0.32\linewidth}
\centerline{\includegraphics[width=0.9\linewidth]{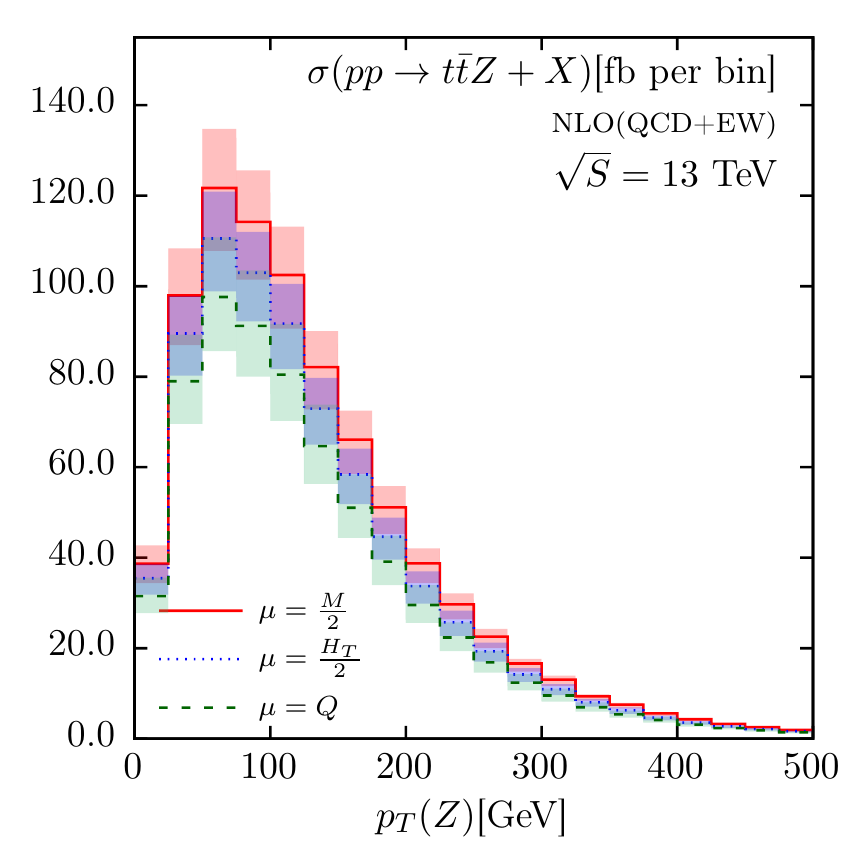}}
\end{minipage}
\hfill
\begin{minipage}{0.32\linewidth}
\centerline{\includegraphics[width=0.9\linewidth]{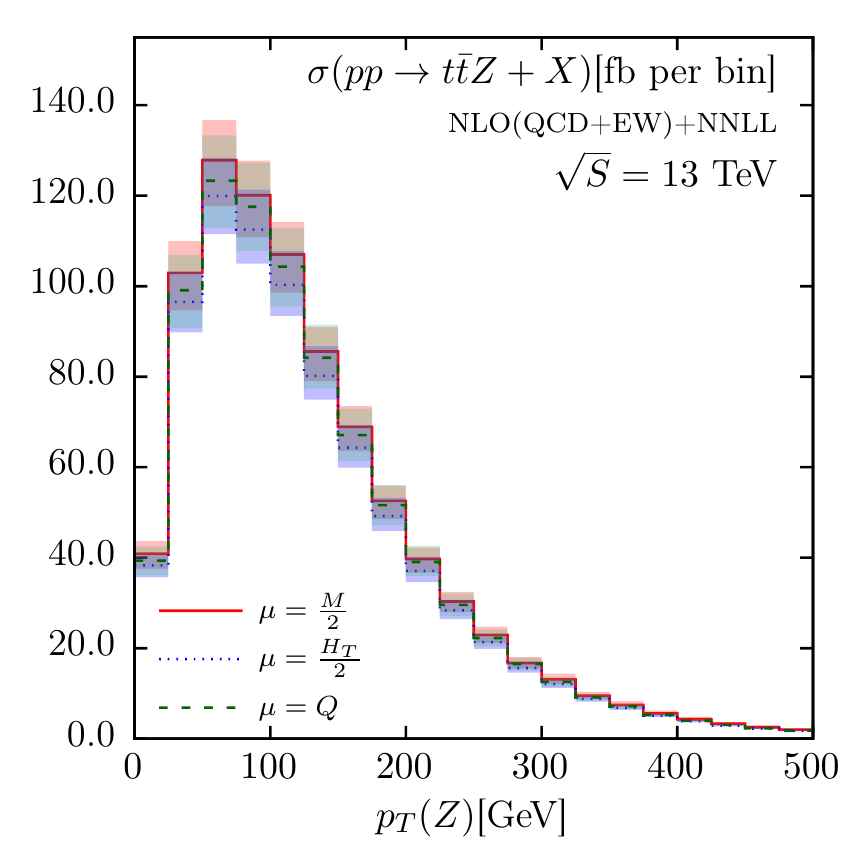}}
\end{minipage}
\hfill
\begin{minipage}{0.32\linewidth}
\centerline{\includegraphics[width=0.9\linewidth]{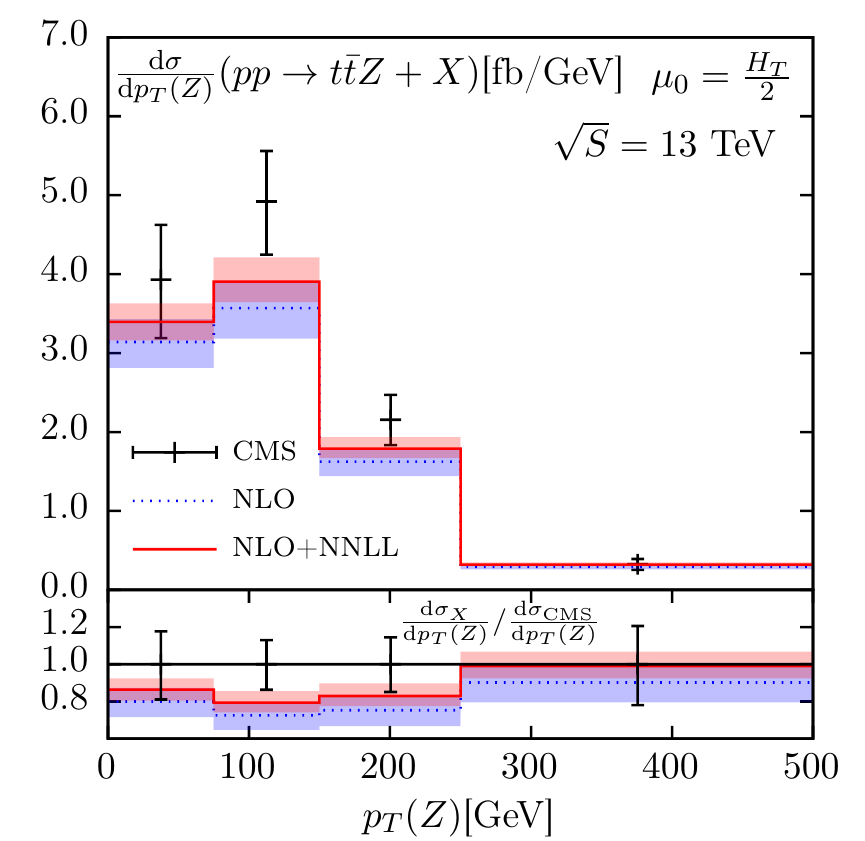}}
\end{minipage}
\caption[]{Dependence of the transverse momentum distribution of the $Z$ boson in $\ttbz$ production at the NLO(QCD+EW) accuracy (left) and  NLO(QCD+EW)+NNLL accuracy (center) on the scale choice. (Right) comparison of the NLO(QCD+EW) and NLO(QCD+EW)+NNLL predictions at $\mu_0=H_T/2$ with data~\cite{CMS:2019nos}.}
\label{fig:ptdistrib}
\end{figure}

Resummation also leads to an improvement of theoretical predictions for the invariant mass distributions~\cite{Kulesza:2017ukk} and transverse momentum ($p_T$) distributions of the EW boson.  In Fig.\ \ref{fig:ptdistrib} we show the $p_T (Z)$ distributions for the $\ttbz$ production. The predictions also include the EW corrections \cite{Frixione:2014qaa}, added~\cite{deFlorian:2016spz} to the NLO and NLO+NNLL results in QCD. By comparing  the left and center plots,  it is clear that the spread in the NLO  predictions due to scale variation is reduced greatly if the NNLL corrections are included in the predictions. In the right plot in Fig.\ \ref{fig:ptdistrib} we compare the NLO(QCD+EW)+NNLL distribution, calculated with  $\mu_0=H_T/2$, to the recent measurement of $p_T(Z)$ by the CMS collaboration~\cite{CMS:2019nos} . The NNLL corrections increase the NLO predictions by about 10\% and bring the theory predictions in full agreement with data.


{\bf Acknowledgments:}
The authors acknowledge financial support by the DFG Grant KU 3103/1 (AK, DS), NCN grant No.\ 2017/27/B/ST2/02755 (LM, TS), and by the DAAD with funds from the BMBF and the European Union (FP7-PEOPLE-2013-COFUND - grant agreement n$^\circ$ 605728) (VT).


\end{document}